\documentclass{elsart}
\usepackage{epsfig}
\usepackage{amsmath}
\usepackage{amssymb}
\begin{document}
\begin{frontmatter}
\title{Wave functions and Bohmian trajectories
in interference phenomena}
\author{\'{E}milie Guay\thanksref{guay}},
\thanks[guay]{Present address: D\'{e}partement de physique,
Universit\'{e} Laval, Qu\'{e}bec, Qc.\ Canada G1K~7P4.}
\ead{emilie.guay.1@ulaval.ca}
\author{Louis Marchildon\corauthref{march}}
\corauth[march]{Corresponding author.}
\ead{marchild@uqtr.ca}
\address{ D\'{e}partement de physique,
Universit\'{e} du Qu\'{e}bec,
Trois-Rivi\`{e}res, Qc.\ Canada G9A~5H7}
\begin{abstract}
In the last few years the hydrodynamic
formulation of quantum mechanics, equivalent to
the Bohmian equations of motion, has been used
to obtain numerical solutions of the Schr\"{o}dinger
equation.  Problems, however, have been experienced
near wave function nodes (or low probability regions).
Here we attempt to compute wave functions and
Bohmian trajectories for the interference of one
particle or of two identical particles.  It turns
out that the large number of nodes (i.e.\ interference minima)
makes the hydrodynamic equations impractical,
whereas a more straightforward solution of the
Schr\"{o}dinger equation gives very good results.
\end{abstract}
\end{frontmatter}
\section{Introduction}
Bohmian trajectories were first proposed in an
attempt to restore determinism in quantum
mechanics~\cite{Bohm_article}.  In Bohm's view,
quantum particles have at every instant 
well-defined positions which, however, can only
be known probabilistically.  The particles follow
deterministic trajectories governed by equations
of motion similar to Newton's, except that
a specific quantum contribution must be added
to the classical potential.  This \emph{quantum
potential} explicitly depends on the particles'
total wave function and is responsible for
characteristic quantum-mechanical effects like
barrier penetration.  All statistical predictions
of quantum mechanics can be obtained through
averages over trajectories~\cite{Holland,Bohm_livre}.

One of the first numerical computations of Bohmian
trajectories was carried out in the context of
two-slit interference~\cite{Philippidis}.
It showed vividly how
one-particle interference effects
can be understood in terms of particle dynamics.
Trajectories associated with two-particle
interference were also shown explicitly to
reproduce standard quantum-mechanical
results~\cite{article}.

Bohmian trajectories can usually be computed
rather straightforwardly if the particles' wave
function is known analytically.  Computation
can also be carried out from a numerical
approximation to the exact wave function.
But this, it turns out, can be viewed
from a different perspective.  In the past
few years, the computation of Bohmian trajectories
has led to a powerful way of numerically integrating
the Schr\"{o}dinger equation.

The method is closely connected with the \emph{hydrodynamic
formulation} of the Schr\"{o}\-dinger equation,
which goes back to the early years of
quantum mechanics~\cite{Madelung,Broglie}.
In this formulation, the evolution of the
wave function is associated with that of a
fluid whose motion can be
obtained through the trajectories of its
elements.  Among the quantum-mechanical
problems that have been adressed in this way
are photodissociation~\cite{Dey,Mayor},
reactive scattering with the Eckart
barrier~\cite{Wyatt,Wyatta},
the quartic double-well potential~\cite{Bittner},
and the harmonic oscillator with quartic
anharmonicity~\cite{Zhao}.  The
method has a number of advantages, in particular
the use of a relatively small number of grid
points and its applicability to higher-dimensional
problems.  It does, however, have difficulties
in dealing with regions where the wave function
vanishes or nearly vanishes.

The purpose of this paper is to investigate
the numerical computation of wave functions and
Bohmian trajectories in the context of particle
interference~\cite{Lam}.  This specifically
quantum-mechanical phenomenon illustrates
perhaps more than any other the properties of
quantum superpositions.  Moreover,
interference minima correspond
to zeros or near-zeros of the wave function,
and therefore make severe tests on
numerical methods.  The hydrodynamic method and
algorithms for its numerical solution will
be reviewed side by side with the method based
upon separating the Schr\"{o}dinger equation
into its real and imaginary parts.  Both will be
used to make detailed computations
of the wave functions and Bohmian
trajectories associated with the interference
of one particle or two identical particles.
Comparison with results obtained through an
exact solution of the Schr\"{o}dinger equation
will show that zeros (or nodes) of the wave function make
the hydrodynamic computation prohibitive in
computer resources, whereas the approach using
the real and imaginary parts of the wave function
yields accurate results in reasonable time.
\section{Wave functions and trajectories}
The Schr\"{o}dinger equation for a system of $n$
particles interacting through a potential $V$ is
given by
\begin{equation}
i\hbar \frac{\partial \psi}{\partial t} 
= - \sum_{i=1}^n \frac{\hbar^2}{2 m_i}\nabla_i^2 \psi +
V(\vec{r}_1, \ldots, \vec{r}_n) \psi , \label{Schrodinger}
\end{equation}
where $m_i$ is the mass of particle $i$ and
$\nabla_i^2$ the Laplacian operator with respect to
that particle's coordinates.
We will focus here on the interference
of one particle or of two identical particles.  There is
then only one mass and, in dimensionless units,
Eq.~(\ref{Schrodinger}) can be written as
\begin{equation}
i \frac{\partial \psi}{\partial t} 
= - \frac{1}{2} \nabla^2 \psi + V(\vec{r}\,) \psi .
\label{Schrodinger1}
\end{equation}
The vector $\vec{r}$ now stands for the position
in configuration space, and the operator $\nabla^2$
for the Laplacian in that space. 
\subsection{Hydrodynamic equations}
The hydrodynamic equations follow from the
Schr\"{o}dinger equation when the wave function
is written in polar form.  They were first used
as a basis for the numerical solution of
Eq.~(\ref{Schrodinger}) a number of years
ago~\cite{Weiner}, but the method has been
substantially improved recently. 

To get the hydrodynamic equations we substitute
$\psi = \sqrt{P} \exp (i S)$ in Eq.~(\ref{Schrodinger1}),
where $S$ and $\sqrt{P}$ are real dimensionless
functions and $\sqrt{P}$ is nonnegative.  Equating
separately the real and imaginary parts
of~(\ref{Schrodinger1}), we obtain
\begin{align}
\frac{\partial P}{\partial t}
&= - \vec{\nabla} \cdot (P \vec{\nabla} S) , \label{P}\\
\frac{\partial S}{\partial t} 
&= - \frac{1}{2} (\vec{\nabla} S)^2 - Q - V(\vec{r}\,) , \label{S}
\end{align}
where $Q$, the quantum potential, is given by
\begin{equation}
Q \equiv - \frac{1}{2} \frac{\nabla^2 \sqrt{P}}
{\sqrt{P}} . \label{Q}
\end{equation}

The Bohmian trajectories of the particles are
defined by writing the following equation for
the velocity in configuration space:
\begin{equation}
\vec{v} = \vec{\nabla} S . \label{momentum}
\end{equation}

We now substitute (\ref{momentum}) in (\ref{P})
and (\ref{S}).  Noting that $\nabla_a v_b = \nabla_b v_a$
($a$ and $b$ are coordinate indices), we find
\begin{align}
\frac{D P}{Dt} &= - P(t) \vec{\nabla}
\cdot \vec{v} \, (t), \label{P2}\\
\frac{D \vec{v}}{Dt} &= - \vec{\nabla} (Q + V) , \label{V1}
\end{align}
where the \emph{Lagrangian derivative} is defined as
\begin{equation}
\frac{D}{Dt} = \frac{\partial}{\partial t} 
+ \vec{v} \, (t)\cdot \vec{\nabla} . \label{dl}
\end{equation}
Eqs.~(\ref{P2}) and~(\ref{V1}), together with
\begin{equation}
\frac{D \vec{r}}{Dt} = \vec{v} \, (t) \label{position},
\end{equation}
must be solved to get the Bohmian trajectories and,
eventually, the wave function.

The numerical solution of Eqs.~(\ref{P2}), (\ref{V1}),
and~(\ref{position}) can be carried out in two
different ways. The first one, called Lagrange's viewpoint,
uses the fact that the Lagrangian derivative represents
the total time derivative with respect to a
moving coordinate system.  Grid points are
then defined which move with
the particles according to Eq.~(\ref{position}).
This method has a number of advantages. First, the Bohmian
trajectories are automatically calculated.  Secondly, as grid
points follow Bohmian trajectories, they remain concentrated in
regions of high probability.  As a consequence, fewer points
are needed since the grid adapts itself throughout
time evolution.  Finally, use of the
Lagrangian derivative gives differential
equations with very few terms.

Euler's viewpoint, in contrast with Lagrange's,
uses a fixed grid.  Eqs.~(\ref{P2}) and~(\ref{V1})
are solved with (\ref{dl}) substituted into them.
That method may be more flexible for the purpose
of computing spatial derivatives.
\subsection{Schr\"{o}dinger equation}
Several methods for the numerical solution
of the Schr\"{o}dinger equation in its original
form were developed over the
years~\cite{Mazur,Goldberg,Heller,Feit}.
Heller's approach~\cite{Heller}, in particular,
was used for the analysis of atom diffraction
by surfaces through computation of Bohmian
trajectories~\cite{Sanz1,Sanz2}.  Here we shall
separate the Schr\"{o}\-dinger equation into its real
and imaginary parts~\cite{Mazur}
by substituting $\psi = \psi_R + i \psi_I$
in~(\ref{Schrodinger1}).  We get
\begin{align}
\frac{\partial \psi_R}{\partial t} 
&= - \frac{1}{2} \nabla^2 \psi_I + V \psi_I , \label{psiR}\\
\frac{\partial \psi_I}{\partial t} 
&= \frac{1}{2} \nabla^2 \psi_R - V \psi_R . \label{psiI}
\end{align}
Starting with the value of $\psi$ at a given
time, these equations are to be solved over an interval
of time.  Bohmian trajectories are then computed
using~(\ref{momentum}) or, explicitly,
\begin{equation}
\frac{d\vec{r} \, (t)}{dt} = \vec{v} \, (t) 
= \vec{\nabla} \left\{\arctan \left(
\frac{\psi_I}{\psi_R} \right) \right\} .
\label{btraj}
\end{equation}
\section{Numerical methods}
In this section, we address the problems of appropriately
evaluating spatial derivatives and carrying out
time integrations, in each of the two schemes considered.
\subsection{Hydrodynamic equations}
In the Lagrangian viewpoint, one needs to compute spatial
derivatives on a grid that changes at each time step.
We use the \emph{moving weighted least squares method}
proposed in Ref.~\cite{Wyatt}.  It consists in fitting a
series of polynomials to values of functions
in a neighborhood of the point were the derivative is to
be evaluated.  Provided that the neighborhood is
small enough, the function to be
differentiated should be well represented by low-order
polynomials.  Function derivatives will then be given
by the coefficients of the polynomials.  This method
can be adapted to almost any point distribution.

To be more specific, suppose we want to
evaluate derivatives of a function $f$
at a point $\vec{r}_0$.  We first write $f$ as a finite series
of polynomials around $\vec{r}_0$, that is,
\begin{equation}
f(\vec{r} \, ) = \sum_{s=1}^{M} a_s 
p_s(\vec{r} - \vec{r}_0). \label{serief}
\end{equation}
We now use the values of $f$ at $N_b$
neighboring points $\vec{r}_n$ of
$\vec{r}_0$, and find the coefficients $a_s$ by minimizing
the following expression:
\begin{equation}
\chi^2 = \sum_{n=1}^{N_b} \left[\frac{f(\vec{r}_n) 
- \sum_{s=1}^{M} a_s p_s(\vec{r}_n
- \vec{r}_0)}{\sigma_n} \right]^2 . \label{chi_carre}
\end{equation}
Introducing the rectangular matrix
\begin{equation}
A_{ns}= \frac{p_s(\vec{r}_n -\vec{r}_0)}{\sigma_n} 
\end{equation}
and the vector
\begin{equation}
b_n = \frac{f(\vec{r}_n)}{\sigma_n},
\end{equation}
we find that $\chi^2$ is minimized if
\begin{equation}
\mathbf{A}^{T} \cdot \vec{b} 
= \mathbf{A}^{T} \cdot \mathbf{A} \cdot \vec{a}.
\label{normal}
\end{equation}
This is called the \emph{normal equation}~\cite{NumRecipes}.
The weighted least squares approximation is implemented
by solving~(\ref{normal}) for the unknown $\vec{a}$.
Note that $\mathbf{A}^{T} \cdot \mathbf{A}$ is a square
matrix.  Once the $a_s$ are known, derivatives of $f$
can be evaluated from Eq.~(\ref{serief}).
The standard error $\sigma_n$ is determined by
assigning larger weights to closer points,
using for instance a Gaussian distribution around $\vec{r}_0$.
Further details on the use of the
weighted least square method in connection with the
hydrodynamic equations can be found
in Refs.~\cite{Wyatta,Bittner}.

Time integration in Lagrange's viewpoint is based
on the following discretization of the Lagrangian
derivative:
\begin{equation}
\frac{Df(t)}{Dt} \rightarrow 
\frac{f(t + \Delta t) - f(t)}{\Delta t}.
\end{equation}
Here $f$ is evaluated on the moving grid,
whose points follow particle trajectories.
In Euler's viewpoint we have
\begin{equation}
\frac{\partial f(t)}{\partial t} \rightarrow
\frac{f(t + \Delta t) - f(t)}{\Delta t} ,
\end{equation}
where $f$ is now evaluated on a fixed grid.

The spatial derivatives turn out to be smoother
if we make the transformation $P = \exp (2g)$.
Making use of Eqs.~(\ref{P2}), (\ref{V1}),
and~(\ref{position}), we find in Lagrange's
viewpoint
\begin{align}
\vec{r}_n(t + \Delta t) 
&= \vec{r}_n(t) + \Delta t \,
\vec{v}_n(t), \label{position1} \\
\vec{v}_n(t + \Delta t)
&= \vec{v}_n(t) - \Delta t \, \vec{\nabla}(Q_n + V_n),
\label{vitesse1} \\
g_n(t + \Delta t) 
&= g_n(t) - \frac{1}{2} \Delta t \,
\vec{\nabla} \cdot \vec{v}_n ,
\label{lnDensite}
\end{align}
where
\begin{equation}
Q_n = -\frac{1}{2} \frac{\nabla^2 \sqrt{P_n}}{\sqrt{P_n}}
= - \frac{1}{2}\left\{ \left( \vec{\nabla} g_n \right)^2
+ \nabla^2 g_n \right\} . \label{potqi}
\end{equation}
Here $f_n(t)$, for instance, stands for the value
of the function $f$ at the grid point $n$ at time $t$,
and $\vec{\nabla} g_n \equiv (\vec{\nabla} g)_n$.

In Euler's viewpoint, the following equations have
to be solved, together with~(\ref{potqi}):
\begin{align}
\vec{v}_n(t + \Delta t) 
&= \vec{v}_n(t) - \Delta t \,
\vec{\nabla}(Q_n + V_n) - \Delta t
\{\vec{v}_n(t) \cdot \vec{\nabla} \} \vec{v}_n(t), \label{vi} \\
g_n(t + \Delta t) 
&= g_n(t) - \frac{1}{2} \Delta t \,
\vec{\nabla} \cdot \vec{v}_n 
- \Delta t \, \vec{v}_n \cdot \vec{\nabla}g_n . \label{gi}
\end{align}
It should be pointed out that a higher-order
scheme like the conventional Runge-Kutta method
cannot be used for time integration with the weighted
least squares method, since the functional form
of the spatial derivatives changes at each time step.
\subsection{Schr\"{o}dinger equation}
The numerical solution of Eqs.~(\ref{psiR})
and~(\ref{psiI}) requires discrete approximations
to the second-order spatial derivatives
of $\psi_R$ and $\psi_I$.  Let $\Delta$ denote
the grid spacing.  We use the following approximation,
which neglects terms of order $\Delta^4$ and
higher:
\begin{align}
\frac{\partial^2 f}{\partial x^2} 
&= \frac{1}{12 \Delta^2} \left[-30 f(x) 
+ 16\{f(x + \Delta) + f(x - \Delta)\} \right. \notag\\
&\qquad\qquad \ \left. \mbox{} - \{f(x +
2\Delta) + f(x - 2\Delta)\} \right]. \label{d2fa}
\end{align}
This formula cannot be used near the
grid boundaries, where $f(x \pm \Delta)$ and
$f(x \pm 2\Delta)$ may not be defined.  In this
case we write, for instance,
\begin{align}
\frac{\partial^2 f}{\partial x^2} 
&= \frac{1}{12 \Delta^2} \left[45 f(x) 
- 154 f(x+\Delta) + 214 f(x + 2\Delta) \right. \notag\\
&\qquad\qquad \left. \mbox{} - 156 f(x + 3 \Delta) 
+ 61 f(x + 4 \Delta) -10 f(x + 5 \Delta) \right] ,
\label{d2fb} \\ \displaybreak[0]
\frac{\partial^2 f}{\partial x^2} 
&= \frac{1}{12 \Delta^2} \left[- 15 f(x)
+ 10 f(x - \Delta)  -4 f(x+\Delta) \right. \notag\\
&\qquad\qquad\left. \mbox{} + 14 f(x + 2\Delta) 
- 6 f(x + 3 \Delta) + f(x + 4 \Delta) \right] ,\label{d2fc}
\end{align}
with similar expressions on the other side of the
grid.

A fourth-order expression for first derivatives 
is also used for the computation of Bohmian
trajectories.

Once the spatial discretization is done,
Eqs.~(\ref{psiR}) and~(\ref{psiI}) read as
\begin{align}
\frac{\partial}{\partial t} \psi_{Rn} 
&= -\frac{1}{2} F_n(\psi_{Im}), \\
\frac{\partial}{\partial t} \psi_{In} 
&= \frac{1}{2} F_n(\psi_{Rm}).
\end{align}
For a grid with $N$ points,
this makes up a system of $2 N$ coupled
first-order differential equations.  From
Eqs.~(\ref{d2fa})--(\ref{d2fc}), one can see that
for a given value of $n$, the index $m$ assumes
up to six different values.

Since oscillations of the real and imaginary parts
of the wave function may be important, the numerical
integration of $\psi_R$ and $\psi_I$ requires an
accurate and stable scheme.  We use the fourth-order
Runge-Kutta method.  Note that we improve on
Ref.~\cite{Mazur} in both the spatial and the
time discretization.
\section{One- and two-particle interference}
An idealized interference setup is shown
in Fig.~1, where parameters
later to be used in wave functions are indicated.
The source S either emits one particle at a time,
which may go through one of the slits and be
detected on the screen.  Or it emits two identical
correlated particles, with identical $x$ momenta
and opposite $y$ momenta, so that if one particle
goes through slit $A$ the other goes through slit $B$.

Let $\psi_A(\vec{r}_i, t)$ and $\psi_B(\vec{r}_i, t)$
be the partial wave functions for particle $i$ going
through slit $A$ or $B$.  Just like the symmetry of the
setup, we assume that $\psi_A$ and $\psi_B$ transform
into each other under reflection through the $x$ axis,
that is,
\begin{equation}
\psi_A(x_i, y_i, t) = \psi_B(x_i, -y_i, t) .
\label{ab}
\end{equation}
The $z$ coordinate is omitted throughout.

For one-particle interference, the global wave
function is given by
\begin{equation}
\Psi_{\mathrm{one}}(\vec{r}, t) 
= \mathcal{N} \left[ \psi_A(\vec{r_1}, t)
+ \psi_B(\vec{r_1}, t) \right] ,
\label{Psi_int1}
\end{equation}
where the configuration space coordinate $\vec{r}$
corresponds to the one-particle coordinate $\vec{r}_1$.
Here $\mathcal{N}$ is a normalization constant.  For
two-particle interference, $\vec{r} = (\vec{r}_1,
\vec{r}_2)$ and we can write 
\begin{equation}
\Psi_{\mathrm{two}}(\vec{r}, t) 
= \mathcal{N} \left[ \psi_A(\vec{r}_1, t) \psi_B(\vec{r}_2, t) 
\pm \psi_B(\vec{r}_1, t) \psi_A(\vec{r}_2, t) \right].
\label{Psi_2int}
\end{equation}
The $+$ sign corresponds to bosons, for which the
global wave function is symmetric under particle
exchange, while the $-$ sign corresponds to fermions,
for which the wave function is antisymmetric.
In the one-particle case the interference pattern
shows up on the screen, whereas in the two-particle case
it is a property of configuration space.

At $t = 0$, the partial wave functions are picked
as plane waves in the $x$-direction, and Gaussian wave
packets in the $y$-direction, centered on the
appropriate slit.  Explicitly,
\begin{equation}
\psi_{A}(\vec{r}_i, t=0) 
= \left( 2\pi \sigma_0^2 \right)^{-1/4}
\exp \left\{ - \frac{(y_i -Y)^2}{4\sigma_0^2}
+ i k_x x_i \right\} ,
\end{equation}
with $\psi_B$ given through~(\ref{ab}).  The time
evolution of such wave functions in free space
(where $V=0$) is known exactly.  It is given by
\begin{equation}
\psi_{A}(\vec{r}_i, t) 
= \left( 2\pi \sigma_t^2 \right)^{-1/4}
\exp \left\{ - \frac{(y_i -Y)^2}{4\sigma_0\sigma_t}
+ i \left[k_x x_i - \frac{k_x^2 t}{2} \right] \right\} ,
\label{psiA}
\end{equation}
where
\begin{equation}
\sigma_{t} = \sigma_{0} \left( 1 + \frac{i t}
{2 \sigma_{0}^{2}} \right) . \label{sigmat}
\end{equation}

For plane waves along $x$, one can show~\cite{article}
that the $x$-coordinate Bohmian trajectory is simply
given by $x(t) = x(0) + k_x t$.  In the numerical
implementation, we therefore concentrate only on
the $y$-coordinates.  One-particle interference thus
reduces to a one-dimensional problem, whereas
two-particle interference is a two-dimensional problem.

We recall that Bohmian trajectories have
been obtained, from exact wave functions, for
one-particle interference in~\cite{Philippidis}
and for two-particle interference in~\cite{article}.
In the numerical computations of wave functions
and trajectories that follow, we let throughout
$Y = 1$, $\sigma_0 = 0.2$, and $k_x = 0.1$.
\section{Results and discussion}
\subsection{Hydrodynamic equations}
One of the main advantages of the Lagrangian viewpoint
is the possibility of concentrating grid points in regions
of high probability.  Accordingly, our first attempts
at solving the
hydrodynamic equations for one-particle interference
used grid points in the immediate neighborhood of the slits
only.  The numerical results obtained with such initial
conditions were very different from what should be expected.
Instead of building up an interference pattern, they
represented essentially
independent Gaussian wave packets emerging
from each slit.  Clearly then, grid points are needed in the
whole region between the slits, even where the probability
of finding a particle is very low.  This is related to the fact
that the development of plateaux and troughs in the quantum
potential responsible for the formation of fringes really
begins around $y=0$~\cite{Philippidis}.  These remarks
point to one of the main problem encountered with the
hydrodynamic equations: wave function
nodes~\cite{Bittner,Zhao,Wyatt5,Wyatt3}.

The reason why nodes or quasinodes of the wave
function are apt to cause problems in the hydrodynamic
approach is apparent from Eq.~(\ref{Q}).  While the denominator of
the quantum potential then nearly vanishes, the numerator
normally does not.  The quantum potential may thus experience
rapid and important variations which challenge approximation
procedures.  Moreover, in the Lagrangian approach the trajectories
tend to group in regions of higher probability, thereby going
away from nodes.  Thus precision is lacking just where the most
important variations of the quantum potential occur.  The problem
is especially acute in our case of one-particle interference.
Nodes then correspond to some of the most interesting parts
of the wave function, namely interference minima.

Since the wave function~(\ref{Psi_int1}) for one-particle
interference nearly vanishes initially at the point
$y=0$, it should help to understand the node problem.
With grid points in the neighborhood of slits only,
no interference pattern is formed and trajectories
go through the node just as if it were absent.
But Bohmian trajectories should never cross, hence
in this case the quantum potential is not properly
calculated.  When points are added near the node, the
quantum potential can be calculated better.  Fig.~2
shows that getting an accurate approximation is
no easy matter.  As expected, the approximation tends
to get better as points are added and more polynomials
are used.  In the Lagrangian approach,
however, the matrix $\mathbf{A}^{T} \cdot \mathbf{A}$
has to be inverted at every point and every time step.
Since the dimension of the matrix is equal
to the number of polynomials, computation
time increases quickly, which makes the method
inefficient.

Moreover, small inaccuracies in the initial quantum
potential cause important effects in later times, as can be
seen in Fig.~3.  Although a good approximation is found
for the quantum potential with 801
grid points between $-4$ and 4 and
fifth-order polynomials, oscillations appear in the quantum
potential and in the velocity as early as $t = 0.01$
(1000 time steps).  These oscillations can be caused by
instabilities in the time integration or by small errors
in the approximation of the spatial derivatives, or both.

The grid used for these tests was uniform.
Although the weighted least square method allows
for nonuniform grids, it would not really help
concentrating points around the initial node.
Other nodes would develop as the interference
pattern builds up, which would
also require additional points. 

Several solutions have been proposed in connection
with the node problem, for example grid
adaptation~\cite{Wyatt2} and a hybrid method consisting
of solving the Schr\"{o}dinger equation near nodes and the
hydrodynamic equations elsewhere~\cite{Wyatt3,Wyatt4}.
In the case of interference, however, nodes are permanent
in time as well as moving in space, so that adaptation
is too expensive.  As far as the Schr\"{o}dinger equation
goes, we shall show in the next section that it is in fact
more efficient than the hydrodynamic equations, and
therefore does not need to be coupled with it.

Instead of using Lagrange's approach, we may try
Euler's.  Since grid points are then fixed,
precision can be controlled easily.  Because matrix
inversion is carried out only at the first
time step, computation times are somewhat shorter.
Yet use of the weighted least square method still
makes this approach inefficient.  As a matter of fact,
the behavior of the quantum potential and velocity
is roughly similar in the Eulerian approach
as in Figs.~2 and~3, since almost no dispersion
of the wave packet has yet occurred at $t=0.01$.
In addition, numerical instabilities arise due to
the terms in Eqs.~(\ref{vi}) and~(\ref{gi}) that
are absent in the Lagrangian approach.  This suggests
that a higher-order scheme is probably needed for
time propagation.

Other ways of approximating derivatives can be used
with Euler's approach and give good results in
reasonable time.  The approximations described
in Sect.~3.2, for example, give very good results
for the initial node.  However, they require more points
than needed in connection with the
Schr\"{o}\-dinger equation,
and the time integration is highly unstable.  In this case
the fourth-order Runge-Kutta scheme could be used.
But again the Schr\"{o}dinger equation looks
more promising, since it involves at most second-order
(instead of third-order) derivatives.

There is no point here to look at two-particle
interference, since the behavior of $\vec{v}$, $Q$, and $P$ is
similar to what was found for one particle, and the
two-particle problem requires a much larger number of points.
\subsection{Schr\"{o}dinger equation}
Figs.~4 and~5 show the
real and imaginary parts of the wave function
for one-particle interference.  Excellent
agreement is found between the numerical solution
of the Schr\"{o}dinger equation (dots) and
the exact value (solid lines).  The grid
spans $y$-values between $-13$ and 13 with
spacing $\Delta$ equal to 0.1, for a total of
261 points.  5000 steps were used to go from $t=0$
to $t=1$.

With the real and imaginary parts of
the wave function in hand, Bohmian trajectories
can be obtained through Eq.~(\ref{btraj}).
Some of these trajectories are shown in dotted
lines in Fig.~6, solid lines representing
trajectories obtained from the exact wave functions.
Again the agreement is excellent.  In general
the grid should cover enough space to avoid boundary
effects.  But with the scheme of Sect.~3.2, second-order
derivatives are computed quickly and, even with 5000 times
steps, the overall computation time (for the wave
function and trajectories) is a few minutes
on a 1.8~GHz Pentium~4 processor.  For comparison,
the computation time of wave functions in the Lagrangian
approach is between one and two orders of magnitude
higher~\cite{note1}.

As mentioned earlier, two-particle interference
here is a two-dimensional problem, and therefore
the grid must be much larger.  We have used a square grid
with $261 \times 261$ points, both $y$-coordinates
going from $-13$ to 13.  The computation time is therefore
much longer.  Bosons wave functions (the $+$ sign in
Eq.~(\ref{Psi_2int})) were used throughout.

Once again results obtained for the real
and imaginary parts of the wave function are
in very good agreement with exact values.
Some Bohmian trajectories, obtained with
\mbox{15\,000} time steps, are shown in
Fig.~7 as dotted lines.  Solid
lines represent trajectories computed with exact
wave functions.  The agreement is usually
very good, with just a small difference showing up
in the upper curve in Fig.~7a.  The main
reason for this is here again that the trajectory
goes through a node of the wave function,
where the velocity varies quickly.  The complete
numerical solution for the wave function and
trajectories is then more demanding.  Similar
behavior was observed for one-particle interference,
but in both cases differences are small, in sharp
contrast with results from the hydrodynamic equations.
\section{Conclusion}
To our knowledge, Bohmian trajectories in the
context of interference through wave packet
spreading have hitherto been calculated
only from exact wave functions.  In this paper, we
have investigated two different methods for the
numerical computation of wave functions.  The
hydrodynamic equations, written either in Lagrange's
or Euler's viewpoint, are sensitive to the
evaluation of derivatives of the quantum potential.
This is especially delicate near wave function
nodes, inevitable in interference problems.
Lagrange's viewpoint is usually attractive because
the grid automatically adapts to regions of high
probability, and because it addresses
higher-dimensional problems with relative ease.
Yet here proper evaluation of derivatives with the
weighted least squares method requires a large
number of points and high-order polynomials, and time
propagation tends to be unstable.  In Euler's viewpoint,
derivatives can be computed more efficiently, and
higher-order schemes can be used for time propagation.
But then it is simpler and more accurate to
use the Schr\"{o}dinger equation directly, which
involves at most second-order derivatives (instead of
the gradient of the quantum potential).  In both
one- and two-particle interference problems, the
direct numerical solution of the Schr\"{o}dinger
equation thus provides an accurate and
relatively quick way of obtaining wave functions and
Bohmian trajectories.
\section*{Acknowledgment}
One of us (EG) would like to thank the Natural Sciences
and Engineering Research Council of Canada for the
award of a postgraduate scholarship.

%
\newpage
\section*{Figure captions}
\noindent Figure~1\@. Two-slit interferometer.

\medskip
\noindent Figure~2\@. Initial quantum potential. (a) 12 neighbors,
401 points, order of polynomials varied; (b) 12 neighbors,
fifth-order polynomials, number of points varied.
  
\medskip
\noindent Figure~3\@. Velocity at $t=0.01$.
Other conditions same as in Fig.~2.

\medskip
\noindent Figure~4\@. Real part of the wave
function at $t = 1$.

\medskip
\noindent Figure~5\@. Imaginary part of the wave
function at $t = 1$.

\medskip
\noindent Figure~6\@. Bohmian trajectories in
one-particle interference.  The $x$-coordinate
is proportional to the $t$-coordinate, since
$x(t) = k_x t$.

\medskip
\noindent Figure~7\@. Bohmian trajectories
associated with pairs of particles.  (a) $y_1(0) = 1$,
$y_2(0) = -0.6$;  (b) $y_1(0) = 1$, $y_2(0) = -1.4$.

\newpage
\noindent \framebox{Figure~1}
\vspace{5cm}

\begin{figure}[htb]
\begin{center}
\epsfbox{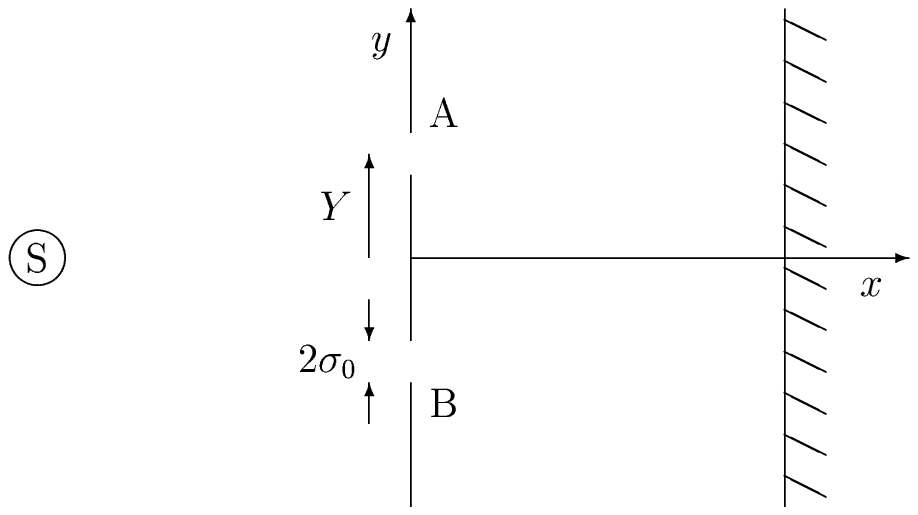}
\end{center}
\end{figure}

\newpage
\noindent \framebox{Figure~2}
\vspace{2cm}

\begin{figure}[htb]
\begin{center}
\epsfbox{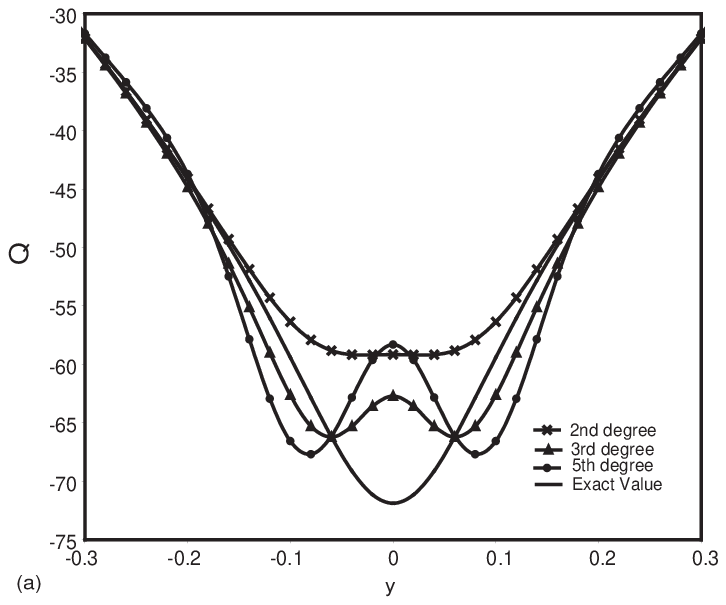}

\vspace{1cm}\hspace{1cm}
\epsfbox{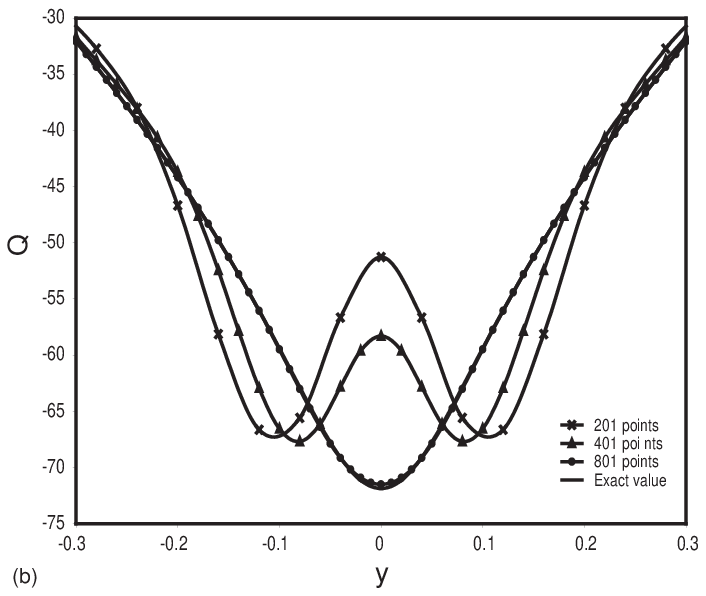}
\end{center}
\end{figure}

\newpage
\noindent \framebox{Figure~3}
\vspace{2cm}

\begin{figure}[htb]
\begin{center}
\epsfbox{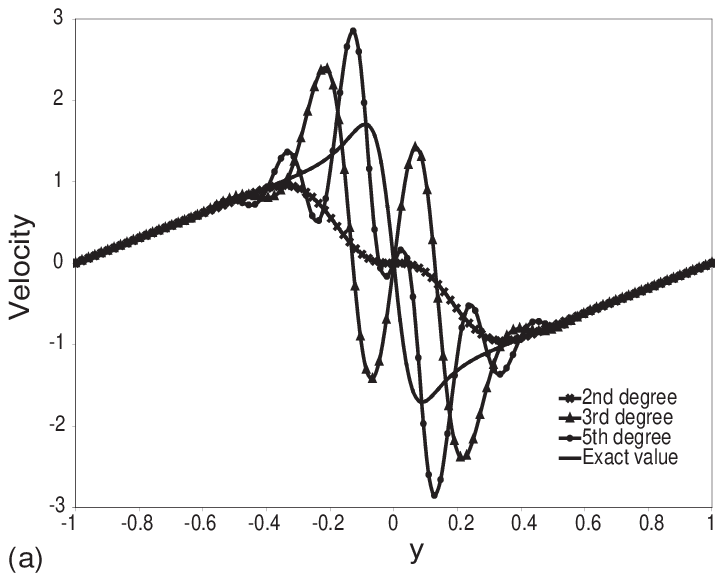}

\vspace{1cm}\hspace{0.7cm}
\epsfbox{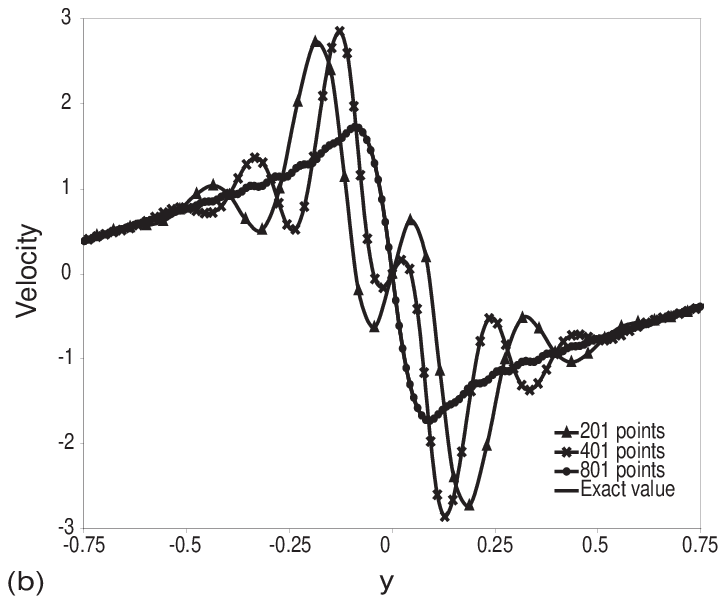}
\end{center}
\end{figure}

\newpage
\noindent \framebox{Figure~4}
\vspace{5cm}

\begin{figure}[htb]
\begin{center}
\epsfbox{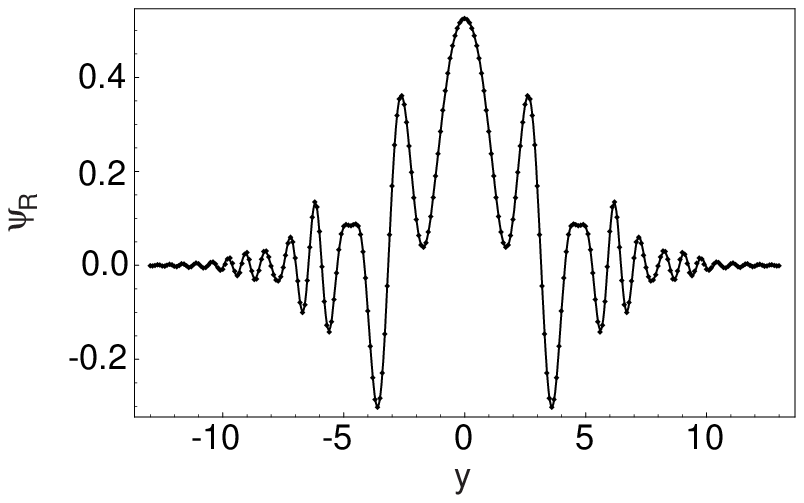}
\end{center}
\end{figure}

\newpage
\noindent \framebox{Figure~5}
\vspace{5cm}

\begin{figure}[htb]
\begin{center}
\epsfbox{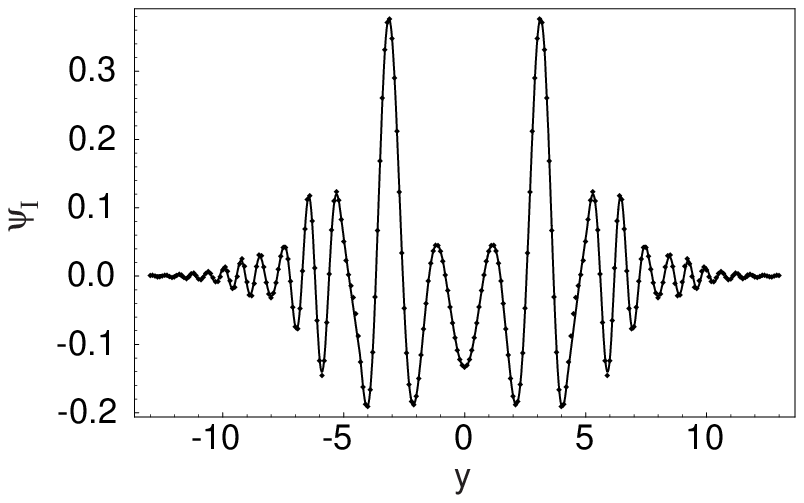}
\end{center}
\end{figure}

\newpage
\noindent \framebox{Figure~6}
\vspace{5cm}

\begin{figure}[htb]
\begin{center}
\epsfbox{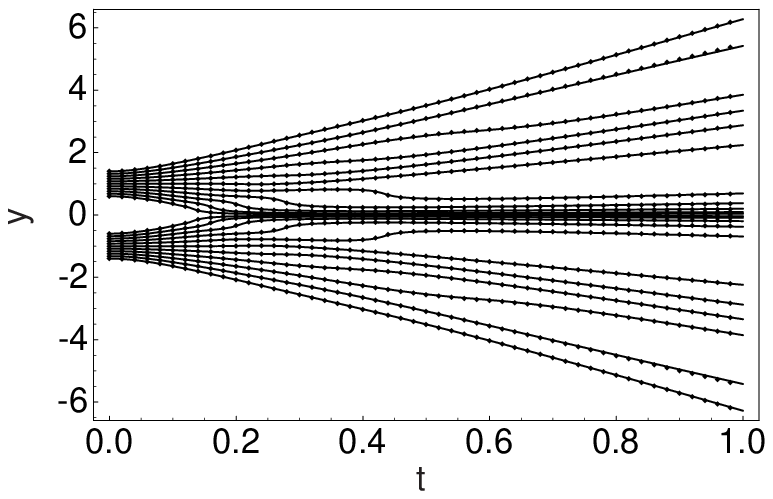}
\end{center}
\end{figure}

\newpage
\noindent \framebox{Figure~7}
\vspace{2cm}

\begin{figure}[htb]
\begin{center}
\epsfbox{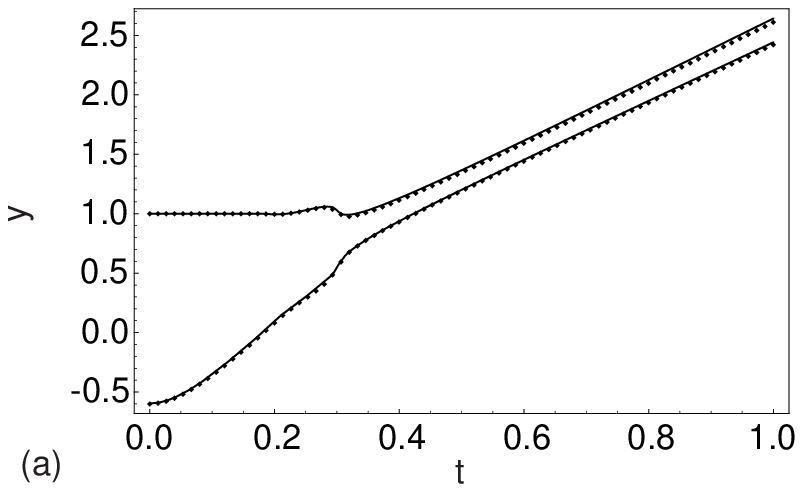}

\vspace{1cm}\hspace{0.1cm}
\epsfbox{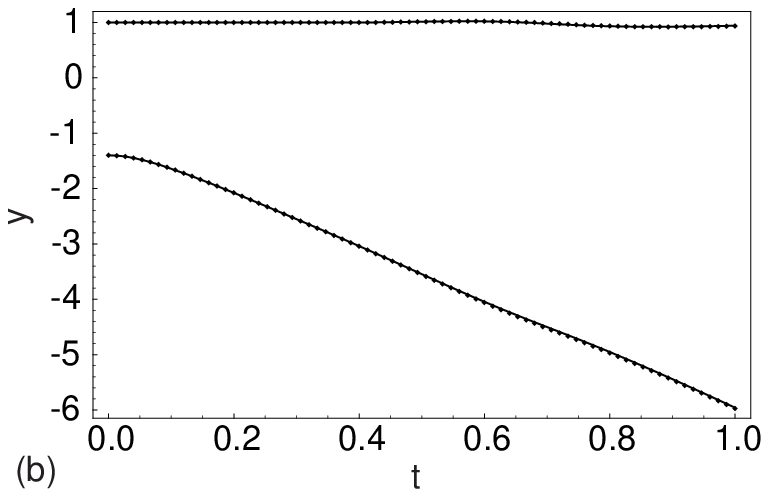}
\end{center}
\end{figure}

\begin{thebibliography}{35}
\bibitem{note1} Our code implementing the Lagrangian
approach can probably be improved for speed.  Yet with
the problem at hand, and for a given number of points
and time steps, the Lagrangian approach is much slower
than the two others considered.
\end{thebibliography}
\end{document}